# ECONOMY OF SCALES IN R&D WITH BLOCK-BUSTERS

**D. Sornette**[1,2]

**Abstract :** Are large scale research programs that include many projects more productive than smaller ones with fewer projects? This problem of economy of scale is relevant for understanding recent mergers in particular in the pharmaceutical industry. We present a quantitative theory based on the characterization of distributions of discounted sales S resulting from new drugs. Assuming that these complementary cumulative distributions have fat tails with approximate power law structure $S^{-\alpha}$, we demonstrate that economy of scales are automatically realized when $\alpha < 1$. Empirical evidence suggests that $\alpha \cong 2/3$ for the pharmaceutical industry.

[1] Institute of Geophysics and Planetary Physics and Department of Earth and Space Sciences, UCLA, Los Angeles, CA 90095-1567

tel: (310) 825 28 63  Fax: (310) 206 3051    email: sornette@moho.ess.ucla.edu

[2] LPMC, CNRS and Université de Nice-Sophia Antipolis, B.P. 71, Parc Valrose

06108 Nice Cedex 2, France

## 1-Introduction

Some industries, such as the pharmaceutical and movie industries, are characterized by the occurrence of "block-busters", i.e. remarkably successful products with exceptional sales much larger than the average. How much exceptional are these block-busters is an important question for firm strategy and the policy of economy of scales (Sornette and Zajdenweber, 1999).

A specific quantification of this observation has been proposed by using the Pareto or power law distribution. Scherer (1965) found that the distribution of sales per drug in the pharmaceutical industry is consistent with a power law distribution as defined in equation (1.1) with exponent $\alpha$ smaller than 0.5. Grabowski and Vernon (1990, 1994) construct a discounted present value per new chemical entity (NCE) and divide the drugs in decile in descending order, leading to a value distribution compatible with a power law distribution of the tail with exponent $\alpha$ approximately equal to 2/3 (Sornette, unpublished). Such a small exponent implies that both the variance and the mean are mathematically infinite and standard economic reasoning and techniques can not apply. Sornette and Zajdenweber (1999) have shown that the distribution of incomes per movie is stable over twenty years and is compatible with a power law distribution with exponent $\alpha$ approximately equal to 1.5. Scherer (1998) revisited a similar problem and investigated empirically the distribution of profits from technological innovations, with the aim to test whether it conforms most closely to the Paretian (power law), log-normal or some other distribution. Scherer (1998) looked at the data in a different way by subtracting the estimated production and marketing costs from sales revenues to obtain a Marshall quasi-rents to R&D investment. He then finds that the distribution of quasi-rents from recent marketed pharmaceutical entities is not a power law but has a thinner tail. This is demonstrated by the downward curvature in a log-log plot (logarithm of the ordinate as a function of the logarithm of the abscissa) This result seems to contradict the older analysis [Scherer, 1965]. However, the procedure of subtracting production and marketing costs introduce a strong bias for intermediate and small sales that are not taken into account.

Indeed, let us call S the cumulative discounted sale from a pharmaceutical innovation. Let us assume that the distribution of sales can be parameterized by the power law

$$P(S) \, dS = \alpha \ (S_{min})^{\alpha} \ S^{-(1+\alpha)} \, dS \qquad \text{where } S_{min} \leq S \, . \qquad (1.1)$$

P(S) dS is the probability for an innovation to produce sales between S and S+dS. We assume a minimum value $S_{min}$ such that the distribution (1) vanished below $S_{min}$. The distribution (1) is normalized. The exponent $\alpha$ controls the frequency of large sales: the smaller $\alpha$ is, the more probable are large sales S relative to smaller ones. Suppose that we look at the distribution of s=S-C, where C is assumed to be the fixed production and marketing costs subtracted to the sales leading to the quasi-rents s. In log-log scales, the distribution of s given by

$$\ln[P(s)] = \text{constant} - (1+\alpha) \ln[s+C] \quad . \tag{1.2}$$

For s>>C, ln[s+C] is indistinguishable from ln[s] and (2) gives a straight line in the log-log plots qualifying a power law. However, when s is not much larger than C, the term ln[s+C] saturates, i.e. instead of - ln[s] growing without bounds, - ln[s+C] saturated to the value - ln[C]. In the log-log representation used for instance by Scherer (1998), this leads to a progressive bent towards the horizontal, which can be misinterpreted as a departure from power law statistics for S. Scherer (1998) analysis thus does not invalidate the power law hypothesis for the distribution of sales.

In the analysis that follows, we start from the hypothesis that the distribution of sales is a power law with exponent $\alpha$ less than one. This case requires a novel and special approach that we develop here. Our main result is that the typical return per product is an increasing function of the total size of the portfolio of products. Our analysis provides an exact benchmark to gauge empirical evidence.

Power law distributions (1.1) are sometimes called ``fractal'' or self-similar (Dubrulle et al., 1997). A power law distribution characterizes the absence of any characteristic size: *independently* of the value of S, the number of realizations larger than $\lambda S$ is a constant factor $\lambda^{-\alpha}$ times the number of realizations larger than S. Suppose S=$10^9$ which occurs with frequency f. Then the frequency of sales equal to $ 2 $10^9$ is $2^{-\alpha}$ f = 0.63 f for $\alpha$=2/3! Take now, S=$10^{10}$ which occurs with frequency f'=$10^{-\alpha}$ f = 0.22 f for $\alpha$=2/3. Then the frequency of $ 2 $10^{10}$ is $2^{-\alpha}$ f' = 0.63 f', i.e. is exactly the same ratio. For any other class of distributions, this ratio of frequencies will depend, not only on the ratio of the values but, also on the absolute value of S. Consider for instance an exponential distribution exp[-S/$S_0$] with characteristic value $S_0$=$10^8$. Sales equal to or larger than S=$7S_0$ occur with a frequency approximately equal to $10^{-3}$. Sales of 2S then occur with frequency $10^{-6}$ giving a ratio of frequency(2S)/frequency(S)=$10^{-6}/10^{-3}$=$10^{-3}$. This ratio is not constant but decreases fast: for instance frequency(4S)/frequency(2S)=$10^{-12}/10^{-6}=10^{-6}$, illustrating the absence of self-similarity in this exponential case.

## 2-Theoretical analysis of the portfolio problem

### 2-1 The $\alpha<1$ power law paradox

Consider a company developing N innovations per year. To simplify the analysis, we discount the total future sales to the value at the inception of the product. The total sale over one year of the company is the sum of its N sales for its N active products for that year:

$$W(N) = S_1 + S_2 + ... + S_N .\qquad(2.1)$$

Before deriving the rigorous analysis in section 2.2, let us provide an heuristic but nevertheless accurate derivation (up to precise numerical pre-factors). For large N, W is approximately

$$W(N) \cong N\, E_N[S]\ ,\qquad(2.2)$$

where $E_N[S]$ is the expectation of the sale size S per product conditioned on the realization of N innovations. The need to impose this condition stems from the fact that the unconditioned expectation is formally infinite for $\alpha<1$ as can be seen by a direct estimation of the integral $\int S\, P(S)\, dS$ from $S_{min}$ to infinity. This divergence reflects the fact that the distribution (1.1) decays so slowly to zero for large S that arbitrarily large values occur with sufficient frequency to draw the expectation to infinity. Of course, this mathematical divergence is not a problem in practice because one always observes a finite sample N with a maximum sale $S_{max}$. The correct expression of the expectation is thus such that it must be performed over all possible values S smaller or equal to $S_{max}(N)$

$$E_N[S] \cong \int_{Smin}^{Smax(N)} S\, P(S)\, dS \qquad(2.3)$$

where the maximum sale $S_{max}(N)$ is a function of the size N of the portfolio of innovations. It is given by

$$N \int_{S_{max}(N)}^{\infty} P(S) \, dS \cong 1 \quad . \tag{2.4}$$

The integral in eq.(2.4) is the probability that one sale is larger than or equal to $S_{max}(N)$. This probability multiplied by N gives the number of sales larger or equal to $S_{max}(N)$. Equating this number with 1 imposes that there is typically only one sale larger than or equal to $S_{max}(N)$. This thus defines the larger sale. Solving (2.4), we get

$$S_{max}(N) \cong S_{min} \, N^{1/\alpha} \quad , \tag{2.5}$$

showing that, as the size N of the portfolio increases, the largest possible sale increases faster than N. Inserting (2.5) into (2.3), we get

$$E_N[S] \cong [\alpha/(1-\alpha)] \, S_{min} \, N^{(1/\alpha - 1)} \quad , \tag{2.6}$$

where we have neglected 1 compared to $N^{(1/\alpha - 1)}$. This is warranted for $\alpha<1$ for which $1/\alpha - 1 > 0$, which ensures that $N^{(1/\alpha - 1)}$ grows with N. Inserting this result (2.6) into the expression (2.2) of the yearly sale W gives

$$W(N) \cong \alpha/(1-\alpha) \, S_{min} \, N^{1/\alpha} \quad \text{for } \alpha \leq 1 \, . \tag{2.7}$$

Note that $S_{max}(N)/W(N)$ is found independent of N: the single maximum sale is a finite fraction of the total cumulative sale! This approximate calculation predicts $S_{max}(N)/W(N) = (1-\alpha)/\alpha = 1/2$ for $\alpha=2/3$ while the exact result is $1-\alpha=1/3$ for $\alpha=2/3$ (Feller, 1971). The main point here is not to be quantitatively precise but to capture correctly the growth of the total sale W(N) faster than N as N grows.

Indeed, the main message of this calculation is that, for $\alpha<1$, W(N) increases with the size N of the portfolio faster than N. The function W(N) has thus the convexity property

$$W(N_A + N_B) > W(N_A) + W(N_B) \ , \tag{2.8}$$

where the equality is recovered for $\alpha \geq 1$.

This result leads to a paradox. Indeed, consider two companies A and B developing respectively $N_A$ and $N_B$ innovations per year and their hypothetical merger C=A+B with $N_A + N_B$ innovations per year. The yearly sales for each company is

$$W_A = S_1 + S_2 + ... + S_{NA} \tag{2.9}$$

$$W_B = S_{NA+1} + S_{NA+2} + ... + S_{NA+NB} \tag{2.10}$$

It is clear that the total sale of C is simply the sum of the sales of A and the sales of B:

$$W_A + W_B = S_1 + S_2 + ... + S_{NA} + S_{NA+1} + S_{NA+2} + ... + S_{NA+NB} \ . \tag{2.11}$$

There thus seems to be little advantage in the merger. However, equation (2.11) is valid for each realization, i.e. year after year. The result (2.8) derives from the fact that we have calculated the typical, in other words, most probable finite size expectations. Thus, the most probable (in a statistical sense) total sale of C is larger than the sum of the most probable total sales of A and of B! In intuitive terms, the mechanism is the following. The total sale $W(N)$ is controlled by the largest sale $S_{max}(N)$ as well as the few largest ones. This is true for portfolios A, B and C. Now, the largest sale $S_{max}(N_A + N_B)$ is equal to either the largest sale of A or the largest sale of B. Let us assume that it is the former. Then in this statistical realization, the portfolio A had a lucky year while portfolio B had a bad year. It may face difficulties or even fill up bankruptcy. For the merged portfolio, the risks are smaller as the most probable sale is larger than the some of the two in the most probable sense. In other words, the most probable largest and total sales are not additive for power law distributions with exponent $\alpha < 1$ This is the essence of the inequality (2.8) that we now put on a rigorous basis.

*2-2 Rigorous analysis of the distributions of aggregate sales*

For S distributed according to the distribution (1.1), the distribution of W(N) converges for large N

to the fully asymmetric (W>0) stable Lévy law $L_\alpha(W(N))$ with exponent $\alpha$ (Gnedenko and Kolmogorov, 1954). The Lévy law has the same power law structure as (1.1) for W large ut presents a smoother behavior for small values. For $\alpha<1$, as for the power law (1.1), its mean and variances are both mathematically infinite. The characteristic scale of the fluctuations is given by a scale parameter C defined by the asymptotic expression of $L_\alpha(W(N))$ for large W:

$$L_\alpha^N(W) = C(N) \, W^{-(1+\alpha)} \qquad \text{for } W \gg 1 . \tag{2.12}$$

There are no analytic expression of stable Lévy law except for a few special values of the exponent $\alpha$. The Lévy law for $\alpha=1$ is the Cauchy (or Lorentz) law. Lévy laws are characterized by their characteristic function which, for $\alpha<1$, reads

$$L_\alpha(k) = e^{-a|k|^\alpha}, \tag{2.13}$$

where a is a constant proportional to the scale parameter C (Gnedenko and Kolmogorov, 1954). The distribution $L_\alpha(W(N))$ can be expressed in a way that makes explicit the dependence on N:

$$L_\alpha^N(W) \, dW = N^{-1/\alpha} \, L_\alpha^1(W/N^{1/\alpha}) \, dW . \tag{2.14}$$

This exact relationship (2.14) means that the distribution of N variables is identical to the Lévy distribution of one variable when W(N) is re-scaled by $N^{1/\alpha}$ and when the distribution $L_\alpha^N(W(N))$ is multiplied by $N^{1/\alpha}$. The expression (2.14) expresses the mathematical property of stability under convolution, that actually defines the Lévy laws.

We can thus write the distribution of yearly sales of the three portfolios A, B and C as follows:

$$P_A(W_A) \, dW_A = N_A^{-1/\alpha} \, L_\alpha^1(W_A/N_A^{1/\alpha}) \, dW_A , \tag{2.15}$$

$$P_B(W_B) \, dW_B = N_B^{-1/\alpha} \, L_\alpha^1(W_B/N_B^{1/\alpha}) \, dW_B , \tag{2.16}$$

$$P_C(W_C) \, dW_C = (N_A + N_B)^{-1/\alpha} L_\alpha^1(W_C / (N_A + N_B)^{1/\alpha}) \, dW_C \, . \qquad (2.17)$$

These expressions formalize rigorously the heuristic derivation leading to (2.8).

To see this, let us consider the following problem, which could be called the bankruptcy problem or the minimum return problem. Let us assume that a portfolio is sustainable if its income $W(N)$ is larger than a fixed factor f times its size N. The factor f embodies the marginal costs associated to each R&D innovation. The total cost per year is thus proportional to the number of innovations. It can also take into account the cost of all the researches that did not lead to any innovation, if the number of successful innovations are proportional to the number of attempts. The important point is that we can quantify the benefits of the merging of the two portfolios A and B by comparing the aggregate costs to the aggregate earnings.

The probability that the portfolio A remains viable is equal to the probability that the total sales $W_A$ be larger than the threshold $f N_A$:

$$p_A = \int_{fN_A}^{+\infty} dW_A \, P_A(W_A) = \int_{x(N_A)}^{+\infty} dx \, L_\alpha^1(x) \, . \qquad (2.18)$$

The second equality in (2.18) is obtained by posing $x = W_A / N_A^{1/\alpha}$ and the lower bound is

$$x(N_A) = f / N_A^{(1/\alpha)-1} \, . \qquad (2.19)$$

The same result holds for portfolio B (resp. C) by replacing $N_A$ by $N_B$ (resp. $N_A + N_B$). Note that the only dependence of $p_A$ in $N_A$ appears in the lower bound (2.19) of the integral. For $\alpha < 1$, the larger $N_A$ is, the smaller is the lower bound (2.19) and the larger is the probability of getting an earning larger than $f N_A$. We thus obtain our fundamental result:

$$\text{if } \alpha < 1, \text{ then } \quad p_C > p_A \text{ and } p_C > p_B \quad \text{for any } f > 0 \, . \qquad (2.20)$$

The probability to obtain an earning per innovation larger than any threshold increases with the total size. If an industry is characterized by very fat tail distributions of sales per product, economy of scale is obtained by growing or merging. From expression (2.7) and inequality (2.8) as well as from the expressions (2.18) and (2.19), we see that the crucial ingredient is the ratio of the *typical* (most probable) sale proportional to $N^{1/\alpha}$ over the size N of the portfolio. Still another way to stress the importance of this ratio is to look at the typical total return per product

$$R = \text{(sales/investment)} \sim N^{1/\alpha - 1} . \qquad (2.21)$$

The larger is the portfolio, the larger is the return per innovation.

It is important to stress that these results emphasizing the comparison between the *typical* (most probable) largest sale per product proportional to $N^{1/\alpha}$ over the size N of the portfolio holds only for $\alpha \leq 1$. For an exponent $\alpha > 1$, the integral defining the expectation converges when the upper bound is put to infinity. The expectation is thus well-defined and the total sale W(N) is simply proportional to N.

## 3-Concluding remarks

Economy of scale and increasing return per innovation has been shown to hold under the simple hypothesis that the distribution of sales per product is a power law with tail exponent $\alpha$ less than one. Its underlying hypotheses are somewhat restrictive: we have neglected any dependence in the research and development of new product as a function of size. Our derivation does not account either for the relevant controlling economic factors, such as level of investment, research organization, spillover, etc. We believe however that this rigorous result provides a useful benchmark as it delineates a simple and well-posed example where economy of scale can be demonstrated unambiguously.

Ackowledgements: I am grateful to W. Comenor, V. Pisarenko and A. Ratcliff for stimulating exchanges.